\documentclass[aps, prb, reprint, twocolumn]{revtex4-1}
\usepackage[T1]{fontenc} 
\usepackage{hyperref}
\hypersetup{colorlinks=true, citecolor=blue, urlcolor=blue, linkcolor=blue}
\usepackage{amsmath}
\usepackage{amsfonts}
\usepackage{ucs}
\usepackage{graphicx,bm,}
\usepackage[caption=false]{subfig}
\usepackage{braket}
\usepackage{breqn}

\makeatletter
\let\cat@comma@active\@empty
\makeatother

\newcommand{\eps}{\epsilon}
\newcommand{\su}{$su(2)$ }
\newcommand{\SU}{SU(2) }
\newcommand{\dn}{\downarrow}
\newcommand{\up}{\uparrow}

\begin{document}
\title{Thermal coupled cluster theory for \SU systems}
\author{Gaurav Harsha}
\affiliation{Department of Physics and Astronomy, Rice University, Houston, Texas 77005, USA}
\author{Yi Xu}
\affiliation{Department of Physics and Astronomy, Rice University, Houston, Texas 77005, USA}
\author{Thomas M. Henderson}
\affiliation{Department of Physics and Astronomy and Department of Chemistry, Rice University, Houston, Texas 77005, USA}
\author{Gustavo E. Scuseria}
\affiliation{Department of Physics and Astronomy and Department of Chemistry, Rice University, Houston, Texas 77005, USA}

\begin{abstract}
  Coupled cluster (CC) has established itself as a powerful theory to study correlated quantum many-body systems. Finite-temperature generalizations of CC theory have attracted considerable interest and have been shown to work as nicely as the ground-state theory. However, most of these recent developments address only fermionic or bosonic systems. The distinct structure of the \su algebra requires the development of a similar thermal CC theory for spin degrees of freedom. In this paper, we provide a formulation of our thermofield-inspired thermal CC for \SU systems. We apply the thermal CC to the Lipkin-Meshkov-Glick system as well as the one-dimensional transverse field Ising model as benchmark applications to highlight the accuracy of thermal CC in the study of finite-temperature phase diagrams in \SU systems.
\end{abstract}

\maketitle

\section{\label{sec:intro}Introduction}
The \su algebra forms the basis for understanding a wide range of phenomena in condensed matter as well as chemical systems. For instance, spin systems, such as the transverse field Ising~\cite{pfeuty_one-dimensional_1970} model and variants of the Heisenberg model, provide one of the most effective ways to study magnetic properties in materials. Hamiltonians for some fermionic systems can also be represented using the $su(2)$ generators, e.g. the Lipkin-Meshkov-Glick~\cite{lipkin_validity_1965, meshkov_validity_1965, glick_validity_1965} and the reduced BCS models.~\cite{bardeen_theory_1957, belyaev_effect_1959, ogle_single-particle_1971}
The computational cost required to get the exact solution to the Schr\"odinger equation (even for the ground state) of interacting quantum many-body systems, which includes the \SU models under consideration, grows exponentially with the size of the system. Significant research has been devoted to finding methods which provide sufficiently accurate solutions to the many-body Schr\"odinger equation at an affordable cost. Prominent examples of such methods for SU(2) systems include various flavors of quantum Monte Carlo (QMC),~\cite{sandvik_quantum_1991, sandvik_stochastic_1999} density matrix renormalization group (DMRG) and tensor-product states,~\cite{white_density-matrix_1993, cirac_renormalization_2009} pseudo fermion functional renormalization group (PFRG),~\cite{kopietz_introduction_2010} cluster-based methods~\cite{li_block-correlated_2004, isaev_hierarchical_2009, fan_cluster_2015, gunst_block_2017, papastathopoulos-katsaros_cluster-based_2021} and coupled cluster (CC) theory.~\cite{crawford_introduction_2000, bartlett_coupled-cluster_2007} Among the deterministic alternatives, DMRG provides highly accurate results for one-dimensional lattices with short-range interactions but is known to struggle with systems in higher dimensions and in the presence of long-range interactions. Contrarily, truncated CC theory in the broken-symmetry basis (e.g., in the on-site basis for spin Hamiltonians, and where symmetries may not be preserved), though not as good as DMRG in one dimension, is reasonably accurate and its performance is not affected by the dimensionality of the lattice. This has led to an extensive application of the CC theory to \SU systems.~\cite{bishop_overview_1991, rosenfeld_extended_1999, rosenfeld_phase_2000, bishop_towards_2006, henderson_quasiparticle_2014, wahlen-strothman_merging_2017, harsha_difference_2018, bishop_frustrated_2019, henderson_correlating_2020, khamoshi_exploring_2021}

The number of accurate methods available for the study of excited states and thermal properties of spin models is markedly smaller than that for ground state. In particular, we have a handful of deterministic methods, such as ancilla DMRG~\cite{verstraete_matrix_2004, feiguin_finite-temperature_2005, czarnik_variational_2016} and thermal generalizations of the PFRG,~\cite{niggemann_frustrated_2021} and a slightly bigger set of stochastic methods such as finite-temperature generalizations of QMC~\cite{rubenstein_finite-temperature_2012, claes_finite-temperature_2017, petras_using_2020, liu_unveiling_2020, shen_finite_2020} and minimally entangled typical thermal states (METTS).~\cite{white_minimally_2009, stoudenmire_minimally_2010}

The development of finite-temperature wave-function methods for electronic structure theory has attracted considerable interest in recent years. Several thermal generalizations of ground-state wave-function methods, specially the CC theory, have been introduced.~\cite{sanyal_thermal_1992,sanyal_systematic_1993,mandal_thermal_1998,mandal_finite-temperature_2003,hermes_finite-temperature_2015, white_time-dependent_2018, hummel_finite_2018, hirata_chapter_2019, harsha_thermofield_2019-1, harsha_thermofield_2019, harsha_wave_2020} This includes our thermofield-inspired coupled cluster~\cite{harsha_thermofield_2019} which constructs a CC approximation to the thermofield double state, a single wave function in an enlarged Hilbert space that provides an exact representation of the thermal density matrix.
Most of these thermal wave function theories, including the thermal CC mentioned above, are tailored to work with bosons or fermions in the grand-canonical ensemble. The difference in the structure of \su and fermion / boson algebras requires further work to extend these thermal wave function theories to spin systems. In this paper, we explore a thermal wave function formalism for \SU systems and construct a thermal CC theory within this framework. To assess its performance, we benchmark our method on the Lipkin-Meshkov-Glick model (referred as the Lipkin model hereafter) and the one-dimensional transverse field Ising model (TFIM).


\section{\label{sec:thermofield}Thermofield dynamics}
Thermofield dynamics~\cite{matsumoto_thermo_1983, semenoff_functional_1983, umezawa_methods_1984, evans_heisenberg_1992} provides a prescription for purification of the finite-temperature ensemble density matrix and constructs a single wave function, often known as the \textit{thermal} or the \textit{thermofield double} state, that can exactly capture the thermal behavior of quantum systems. It does so by working in an enlarged space which comprises the original Hilbert space and a conjugate copy. The ensemble thermal average in the physical space becomes an expectation value over the purified thermal state in the doubled space,
\begin{equation}
  \langle \mathcal{O} \rangle = \mathrm{Tr}\left( e^{-\beta H} \mathcal{O} \right)
  = \frac{\langle \Psi(\beta) \vert \mathcal{O} \vert \Psi (\beta) \rangle}{\langle \Psi (\beta) \vert \Psi (\beta) \rangle},
\end{equation}
where $\beta$ is the inverse temperature and $H$ is the Hamiltonian.
For fermions in the grand-canonical ensemble, we define the thermal state $\vert \Psi (\beta) \rangle$ as
\begin{subequations}
  \begin{align}
    \vert \Psi (\beta) \rangle &= e^{-\beta H / 2} \vert \mathbb{I} \rangle,
    \label{thermal-state}
    \\
    \vert \Psi (0) \rangle &= \vert \mathbb{I} \rangle = \prod_{p} \Big ( 1 + c_p^\dagger \tilde{c}_p^\dagger \Big ) \vert -; - \rangle,
    \label{inf-temp-gcan-thermal-state}
  \end{align}
\end{subequations}
where $c^\dagger_p$ ($\tilde{c}^\dagger_p$) creates a particle in the $p^{\mathrm{th}}$ spin orbital in the physical (auxiliary) space, and the product in Eq.~\ref{inf-temp-gcan-thermal-state} runs over all spin-orbital indices, while the state $\ket{-;-}$  denotes the vacuum for both the physical and conjugate spaces.
The identity state $\vert \mathbb{I} \rangle$ is the exact infinite-temperature thermal state, and derives its name from the infinite-temperature density matrix, which is the identity matrix. The norm of the thermal state gives the partition function.
For \SU spins, we adopt a similar definition for the identity state that was proposed in Refs. \onlinecite{suzuki_thermo_1986, hatsuda_mean_1989, walet_thermal_1990},
\begin{subequations}
  \begin{align}
    \ket{\Psi (0)} &= \ket{\mathbb{I}} = \prod_p \Big (1 + J^+_p \tilde{J}^+_p \Big)  \ket{0},
    \label{inf-temp-spin-thermal-state}
    \\
    \ket{0} &= \ket{\dn \dn \dn \cdots; \dn \dn \dn \cdots},
    \label{all-dn-state}
  \end{align}
\end{subequations}
where $J^+_p$ ($\tilde{J}^+_p$) is the conventional spin-$1/2$ ladder operator for the $p^{\mathrm{th}}$ physical (auxiliary) spin and the state $\ket{0}$ describes all the physical and auxiliary spins, which are written respectively to the left and right of the semicolon in Eq.~\ref{all-dn-state}, pointing downwards.

By definition, the thermal state obeys the imaginary-time evolution equation,
\begin{equation}
  \frac{\partial}{\partial \beta} \vert \Psi (\beta) \rangle = -\frac{1}{2} H \vert \Psi(\beta) \rangle,
  \label{imag-time-evol-eq}
\end{equation}
which can be integrated from $\beta=0$, where $\ket{\mathbb{I}}$ is the exact thermal state, to the desired value of $\beta$. Exact integration, however, scales exponentially with the system size and approximations to this imaginary-time evolution are required.
We have explored similar theory to study both the canonical and grand-canonical ensemble thermal properties of correlated fermionic systems using thermal generalizations of configuration interaction and coupled cluster theory, and refer the reader to Refs.~\onlinecite{harsha_thermofield_2019-1, harsha_thermofield_2019, harsha_wave_2020} and references therein for further details on thermofield theory.

\subsection{Mean-field theory}
The simplest approximation to construct the thermal state is the mean-field approach, where an effective one-body Hamiltonian of the form $H_0 = \sum_p \eps_p J^z_p$ is used to evolve the thermal state and results in
\begin{align}
  \vert \Phi (\beta) \rangle &= e^{-\beta H_0 / 2} \vert \mathbb{I} \rangle,
  \nonumber
  \\
  &= \prod_p \Big(e^{\beta \eps_p / 4} + e^{-\beta \eps_p / 4} J^+_p \tilde{J}^+_p\Big) \ket{0},
\end{align}
which we can normalize into a spin-BCS form,
\begin{equation}
  \ket{0 (\beta)} = \prod_p \Big( u_p + v_p J^+_p \tilde{J}^+_p \Big) \ket{0},
\end{equation}
where $u_p = 1/\sqrt{1 + e^{-\beta \eps_p}}$ and $v_p = \sqrt{1 - u_p^2}$. The BCS parameters, $u_p$ and $v_p$ can also be found by minimizing the mean-field free energy of the system. In the following discussion and results, we will not consider such a reference optimization.

\subsection{Correlated theory}
The mean-field thermal state serves as a reference point to build correlated approximations to the thermal state. Typically, a wave operator $\Omega (\beta)$ is used to build a configuration interaction (CI)- or CC-like expansion of the wave function,
\begin{equation}
  \ket{\Psi (\beta)} \simeq \Omega(\beta) \, \ket{\Phi (\beta)}.
\end{equation}
This form of the correlated thermal state is reminiscent of the interaction picture approach. In most ground-state correlated wave-function theories, it is convenient to express the wave operator $\Omega$ as excitations on the mean-field reference.
For spin systems, this is achieved by transforming the problem to a new \su basis in which the mean-field is a vector product of down quasispins at each site. Then, $\Omega$ can simply be built out of the transformed $J^+$ ladder operators.
Similarly, for thermal wave function theories, we use a canonical transformation that rotates the operator basis in such a way that in the new basis, the thermal mean-field state $\ket{0(\beta)}$ has the same form as $\ket{0}$ in Eq.~\ref{all-dn-state}, i.e., with all the physical and auxiliary quasi spins pointing downwards.
At each lattice site, the fifteen generators,
\begin{equation*}
  J^{\mu}, \tilde{J}^{\nu}, 
  J^\mu \otimes \tilde{J}^\nu, \quad \forall \quad \mu, \nu \in \{\pm, z\}
\end{equation*}
collectively span the $su(4)$ algebra.
The thermal canonical transformation that we seek is a basis rotation in this $su(4)$ algebra. It was first proposed by Suzuki et. al.,~\cite{suzuki_thermo_1986} and is defined as
\begin{subequations}
  \label{su2_transformation}
  \begin{align}
    S^{\pm}_p (\beta) &=
      u_p J^{\pm}_p + 2 v_p J^z_p \tilde{J}^{\mp}_p,
    \\
    \tilde{S}^{\pm}_p (\beta) &=
      u_p \tilde{J}^{\pm}_p + 2 v_p \tilde{J}^z_p J^{\mp}_p,
    \\
    S^z_p (\beta) &=
      u^2_p J^z_p - v^2_p \tilde{J}^z_p - u_p v_p \left(
        J^+_p \tilde{J}^+_p + J^-_p \tilde{J}^-_p
      \right),
    \\
    \tilde{S}^z_p (\beta) &=
      u^2_p \tilde{J}^z_p - v^2_p J^z_p - u_p v_p \left(
        J^+_p \tilde{J}^+_p + J^-_p \tilde{J}^-_p
      \right),
  \end{align}
\end{subequations}
where the coefficients $u_p$ and $v_p$ are same as the ones discussed above. The new $S^-$ and $\tilde{S}^-$ operators annihilate the mean-field reference, i.e.,
\begin{equation}
  S^-_p (\beta) \ket{0(\beta)} = 0 = \tilde{S}^-_p (\beta) \ket{0(\beta)}.
\end{equation}
The transformation in Eq.~\ref{su2_transformation} can, in fact, be derived by realizing that
\begin{subequations}
  \begin{align}
    & e^{-\beta H_0 / 2} S^-_p (0) e^{\beta H_0 / 2} \ket{0 (\beta)} = 0,
    \\
    \Rightarrow &\quad
    S^-_p(\beta) = e^{-\beta H_0 / 2} S^-_p (0) e^{\beta H_0 / 2},
  \end{align}
\end{subequations}
where $S^-_p (0) = (J^-_p + 2 J^z_p \tilde{J}^+_p) / \sqrt{2}$.
The inverse transformation of Eq.~\ref{su2_transformation} can be obtained by swapping $J$ with $S$, and taking $v_p \rightarrow -v_p$.
For the sake of brevity, in the remainder of this manuscript, we will use the label $S$ for thermal and $J$ for zero-temperature \su operators, and drop the explicit $\beta$-dependence.

One can also directly envision the basis rotation in Eq.~\ref{su2_transformation} as a non-linear canonical transformation of the \su algebra. In our implementation, we prefer to work with the $su(4)$ representation.
We use the symbolic algebraic manipulator \textit{drudge}~\cite{zhao_symbolic_2018} to encode the $su(4)$ commutation relations and perform the necessary operator algebra to obtain the expressions for the equations discussed below.

\section{\label{sec:cc-theory}Coupled cluster theory}
In coupled cluster theory, we parameterize the wave function using the exponential of an excitation operator acting on a mean-field reference state. For the thermal state, we get
\begin{equation}
  \ket{\Psi(\beta)} = e^{T(\beta)} \ket{\Phi (\beta)},
  \label{cc-ansatz}
\end{equation}
where $T(\beta)$ creates excitations on the thermal mean-field reference $\ket{\Phi(\beta)}$.
The exponential form generally assures that the computed properties are size extensive and size consistent provided the reference state has these properties to begin with.
For all practical applications, the cluster operator $T$ is truncated to a finite order in excitation rank, e.g., CC truncated to single and double excitation is called CCSD.
Due to the non-linear nature of the thermal transformation in Eq.~\ref{su2_transformation}, the Hamiltonian becomes quartic in \su generators, although it remains quadratic in the $su(4)$ generators. Therefore, we choose a cluster operator which is quadratic in terms of the $su(4)$ generators in order to capture the exact finite-temperature behavior in the simplest two-site systems.
Although symmetries of the system under consideration can be used to simplify its structure, the most general form of the cluster operator $T(\beta)$, with single and double excitations, is
\begin{subequations}
  \label{ccsd-ansatz}
  \begin{align}
    T (\beta) &= t_0 + T_1 + T_2,
    \\
    T_1 &= \sum_p t_p S^+_p + \sum_p \tilde{t}_p \tilde{S}^+_p + \sum_p \alpha_p Y^{++}_p,
    \\
    T_2 &= \frac{1}{2} \sum_{pq} \left(
      t_{pq} S^+_p S^+_q + \tilde{t}_{pq} \tilde{S}^+_p \tilde{S}^+_q
      + \alpha_{pq} Y^{++}_p Y^{++}_q
    \right) \nonumber\\
    & \quad + \sum_{pq} m_{pq} S^+_p \tilde{S}^+_q,
  \end{align}
\end{subequations}
where we define $Y^{\mu\nu}_p = S^{\mu}_p \otimes \tilde{S}^\nu_p$.
The scalar parameter $t_0$ keeps track of the norm of the thermal CC state (i.e., the partition function).
For brevity, we will write $T = \sum_\mu t_\mu \tau_\mu$, where $t_\mu$ and $\tau_\mu$ are compact notations for amplitudes and operators, respectively.
Substituting the CC ansatz (Eq.~\ref{cc-ansatz}) into the imaginary-time evolution equation (Eq.~\ref{imag-time-evol-eq}), we get
\begin{equation}
  \left(e^{-T}\frac{\partial}{\partial \beta} e^{T}\right) \ket{\Phi (\beta)}
  =
  -\frac{1}{2} \left(\bar{H} - H_0\right) \ket{\Phi (\beta)},
  \label{cc-subst-evol-eq}
\end{equation}
where the similarity transformed Hamiltonian, $\bar{H} = e^{-T} H e^{T}$, can be expanded using the Baker-Campbell-Hausdorff (BCH) expansion. While the BCH expansion truncates at the fourth order for a general fermionic Hamiltonian, due to the nontrivial nature of the transformation in Eq.~\ref{su2_transformation}, it truncates at eighth order for \SU Hamiltonians.

The cluster operator $T$ is constructed from the ladder operators in a $\beta$-dependent basis. In general,
\begin{equation}
  \left[ \frac{\partial \tau_\mu}{\partial \beta}, T\right] \neq 0,
\end{equation}
for $\tau_\mu \in \{S^\pm, \tilde{S}^\pm, S^z, \tilde{S}^z, \ldots\}$. Therefore, the similarity transformation of the $\beta$derivative in the left hand side of Eq.~\ref{cc-subst-evol-eq} should be performed using the Wilcox identity~\cite{wilcox_exponential_1967} (see Ref.~\onlinecite{harsha_thermofield_2019-1} for details) and gives
\begin{equation}
  e^{-T} \frac{\partial}{\partial \beta} e^{T}
  =
  \sum_\mu \frac{\partial t_\mu}{\partial \beta} \tau_\mu + D,
\end{equation}
where $D$ represents the contributions from the derivative of the operator part of $T$, and is given by
\begin{subequations}
  \begin{align}
    D &= \sum_\mu t_\mu \bar{\tau}_\mu,
    \\
    \bar{\tau}_{\mu}
    &=
    \frac{\partial \hat{\tau}_\mu}{\partial \beta}
    + \frac{1}{2!} \left[
      \frac{\partial \hat{\tau}_\mu}{\partial \beta}, T
    \right]
    + \frac{1}{3!} \left[
      \left[
      \frac{\partial \hat{\tau}_\mu}{\partial \beta}, T
      \right],
      T
    \right]
    + \ldots
  \end{align}
\end{subequations}
After these manipulations, we arrive at the imaginary-time evolution equation for the amplitudes,
\begin{equation}
  \sum_\mu \frac{\partial t_\mu}{\partial \beta} \tau_\mu \ket{\Phi (\beta)}
  =
  \Big[
    -\frac{1}{2} \left(
      \bar{H} - H_0
    \right)
    - D
  \Big] \ket{\Phi (\beta)},
\end{equation}
which can be projected against various subspaces to yield the evolution equations for the CC parameters $\{t_\mu\}$,
\begin{equation}
  \sum_\mu
  \braket{\tau_\nu^\dagger \tau_\mu} \frac{\partial t_\mu}{\partial \beta}
  =
  \braket{
    \tau_\nu^\dagger
    \Big[
      -\frac{1}{2} \left(
        \bar{H} - H_0
      \right)
      - D
    \Big]
  },
  \label{ivp-equation}
\end{equation}
where the expectation value is calculated over the normalized mean-field thermal state, i.e.,
\begin{equation}
  \braket{X} = \braket{0 (\beta) | X | 0(\beta)} = \frac{
    \braket{\Phi (\beta) | X | \Phi (\beta)}
  }{
    \braket{\Phi (\beta) | \Phi (\beta)}
  }.
\end{equation}
In practice, the excitation operators are orthogonal, so $\braket{\tau_\mu^\dagger \tau_\nu} = \delta_{\mu \nu} \braket{\tau_\nu^\dagger \tau_\nu}$.
The system of first-order differential equations in Eq.~\ref{ivp-equation} can be integrated starting from $\beta = 0$, where the exact initial value for the cluster amplitudes is known ($t_\mu (\beta = 0) = 0$), to the desired inverse temperature.
For all the results discussed in this paper, we use \textit{dopri5},~\cite{dormand_family_1980, hairer_solving_1993} a fourth-order Runge-Kutta algorithm with adaptive grid size (available in SciPy~\cite{2020SciPy-NMeth}), with a tolerance value of $10^{-5}$ to perform the integration.

\section{\label{sec:exp-val}Thermal Properties}
Analogous to the ground-state methods, in approximate finite-temperature wave function theories, we generally have two different ways to evaluate thermal properties: expectation values and free-energy derivatives.
In this section, we will provide a brief overview of these techniques. Further details are discussed in Appendix~\ref{app:thermal-properties}.

\subsection{CC expectation value}
For coupled cluster theory, in the first approach, we compute properties as asymmetric expectation values. For an observable $\mathcal{O}$, we have
\begin{equation}
  \braket{\mathcal{O}} = \braket{0(\beta) | (1 + Z) e^{-T} \mathcal{O} e^{T} | 0(\beta)},
  \label{lin-resp-exp-val}
\end{equation}
where we have a CC approximation for the \textit{ket}, and the \textit{bra}, approximated using CI, is defined as
\begin{equation}
  \bra{\Psi_{CI} (\beta)} = \bra{\Phi (\beta)} (1 + Z) e^{z_0} e^{-T}.
\end{equation}
The scalar $z_0$ tracks the norm of the \textit{bra} thermal state and the CI operator $Z$ is given by
\begin{subequations}
  \begin{align}
    Z &= Z_1 + Z_2,
    \\
    Z_1 &= \sum_p z_p S^-_p + \sum_p \tilde{z}_p \tilde{S}^-_p + \sum_p \gamma_p Y^{--}_p,
    \\
    Z_2 &= \frac{1}{2} \sum_{pq} \left(
      z_{pq} S^-_p S^-_q + \tilde{z}_{pq} \tilde{S}^-_p \tilde{S}^-_q
      + \gamma_{pq} Y^{--}_p Y^{--}_q
    \right) \nonumber\\
    & \quad + \sum_{pq} l_{pq} S^-_p \tilde{S}^-_q.
  \end{align}
\end{subequations}
We can reparametrize the \textit{bra} thermal state as
\begin{equation}
  \bra{\Psi_{CI} (\beta)} = \bra{\Phi (\beta)} (1 + W) e^{w_0},
\end{equation}
such that the $W$ operator has the same structure as $Z$, and the two are related through a disentangled similarity transformation, 
\begin{equation}
  \bra{\Phi (\beta)} (1 + Z) e^{z_0} e^{-T}
  = \bra{\Phi (\beta)} (1 + W) e^{w_0-t_0}.
\end{equation}
The thermal \textit{bra} obeys an imaginary-time evolution equation, similar to the \textit{ket},
\begin{equation}
  \frac{
    \partial \bra{\Psi_{CI} (\beta)}
  }{
    \partial \beta
  } = -\frac{1}{2} \bra{\Psi_{CI} (\beta)} H,
\end{equation} 
from which we can derive the evolution equations for the coefficients in the $W$ (or the $Z$) operator, in the same way as for the CC theory discussed in Sec.~\ref{sec:cc-theory}.

\subsection{Free energy derivatives}
In the second approach, which we also call the $\lambda$-derivative approach, we use Lagrange multipliers to compute finite-temperature properties.
First, we define the $\lambda$-dependent Hamiltonian as $H (\lambda) = H + \lambda \mathcal{O}$, and use it perform the imaginary-time evolution of the thermal state. Then, the property $\mathcal{O}$ can be defined as the $\lambda$ derivative of the free energy $F$, as
\begin{equation}
  \braket{\mathcal{O}} (\lambda) = \frac{\partial F}{\partial \lambda}.
  \label{lambda-exp}
\end{equation}
Usually, we are interested in evaluating $\braket{\mathcal{O}} (\lambda=0)$.
For thermal CC, the partition function and the free energy $F$ are defined as
\begin{subequations}
  \begin{align}
    \mathcal{Z} &= \braket{\Psi_{CI} | \Psi_{CC}}
    = e^{t_0 + z_0} \braket{\Phi (\beta) | \Phi (\beta)},
    \\
    F &= -\frac{1}{\beta} \log \mathcal{Z}.
  \end{align}
\end{subequations}
We can also define the free energy in terms of the integral of internal energy,
\begin{equation}
  F(\beta) = \frac{1}{\beta} \int_0^\beta d\tau E(\tau).
\end{equation}
While constructing the thermal state in the $\lambda$-derivative approach, we should ideally use a mean-field reference that also depends on $\lambda$, i.e., define the partition function as
\begin{equation}
  \frac{\mathcal{Z}}{\mathcal{Z}_0}
  = e^{w_0 + t_0} \braket{
    0 (\beta, \lambda) |
    (1 + W) e^T |
    0 (\beta, \lambda)
  }.
\end{equation}
Comparing with \textit{ab initio} CC theory, a $\lambda$-dependent reference is similar to orbital-optimized linear response CC theory.
Properties calculated using CC expectation values and the $\lambda$-derivative approach, with or without an optimized mean field, are generally different (see Appendix~\ref{app:thermal-properties} for a proof).
However, as the CC approximation becomes accurate, properties from CC expectation and $\lambda$-derivative formalisms will become equivalent. We present results for both these techniques in the following section.

\subsection{Definition of thermal state}
For all the theory and results presented in this paper, we define the \textit{bra} and the \textit{ket} thermal states as
\begin{equation}
  \bra{\Psi (\beta)} = \bra{\mathbb{I}} e^{-\beta H / 2},
  \quad
  \ket{\Psi (\beta)} = e^{-\beta H / 2} \ket{\mathbb{I}}.
\end{equation}
However, as has been discussed in Ref.~\onlinecite{harsha_thermofield_2019-1}, we can also define these thermal states as
\begin{equation}
  \bra{\Psi_\sigma (\beta)} = \bra{\mathbb{I}} e^{-(1 - \sigma)\beta H},
  \quad
  \ket{\Psi_\sigma (\beta)} = e^{-\sigma \beta H} \ket{\mathbb{I}},
\end{equation}
where $0 \leq \sigma \leq 1$, such that the thermal expectation value of an observable $\mathcal{O}$ can be computed as
\begin{subequations}
  \label{sigma-thermal-properties}
  \begin{align}
    \braket{\mathcal{O}} &= \braket{\Psi_\sigma (\beta) | \mathcal{O} | \Psi_\sigma (\beta)},
    \\
    &= \braket{\mathbb{I} | e^{-(1 - \sigma) \beta H} \mathcal{O} e^{-\sigma \beta H}| \mathbb{I}}.
  \end{align}
\end{subequations}
Note that the thermal state defined in Eq.~\ref{thermal-state} is simply $\ket{\Psi_{\sigma=1/2} (\beta)}$.
Using the fact that $\braket{\mathbb{I} | X | \mathbb{I}} = \mathrm{Tr} X$, combined with the cyclic property of a trace, it is easy to show that Eq.~\ref{sigma-thermal-properties} yields the correct ensemble average for the physical observable $\mathcal{O}$.

In the exact theory as well as for mean-field and CI approximations, thermal properties can be computed as symmetric ($\sigma = 1/2$) or asymmetric ($\sigma \neq 1/2$) expectation values. On the other hand, thermal CC expectation values are inherently asymmetric, even for $\sigma = 1/2$, because a simultaneous exponential parametrization of both the \textit{bra} and the \textit{ket} is computationally unfeasible.

\section{Results}
We study the Lipkin model and the transverse field Ising model (TFIM) to assess the performance of our thermal CC theory. As indicated in Eq.~\ref{ccsd-ansatz}, we consider a cluster operator that is quadratic in $su(4)$ generators. For the Lipkin model, we present results for error in internal energy, while we consider both the energetics and properties for TFIM.

\begin{figure*}
  \centering
  \includegraphics[width=0.85\linewidth]{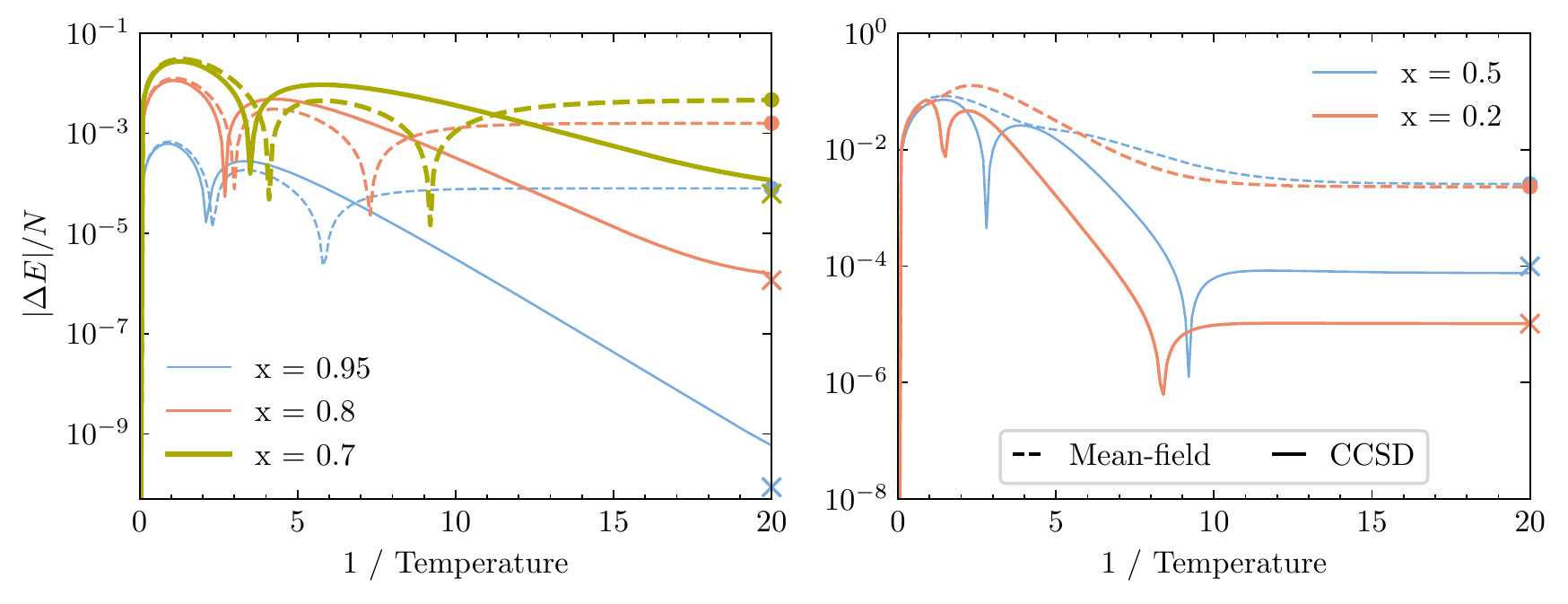}
  \caption{Absolute error in internal energy per site as a function of the inverse temperature for the 32-site Lipkin model in both the weakly interacting regime (left), where the mean-field Hamiltonian $H_0$ preserves the symmetry of the ground state, and strongly interacting regime (right), where the ground-state symmetry is broken at the level of mean field. The plots compare mean-field theory and thermal CCSD against exact results. The colored ``o'' and ``x'' markers on the $y$-axis indicate the corresponding energy error per site for ground-state RHF/UHF and CCSD, respectively.}
  \label{fig:lipkin-error}
\end{figure*}

\subsection{Lipkin-Meshkov-Glick model}
The Lipkin-Meshkov-Glick~\cite{lipkin_validity_1965,meshkov_validity_1965,glick_validity_1965} model describes a closed-shell nucleus with schematic monopole interactions. The system consists of $N$ spins on a lattice in the presence of an external magnetic field in the $z$ direction. Any two spin-up (spin-down) states can flip to spin-down (spin-up) states and lower the energy of the system. The Hamiltonian can be expressed as
\begin{equation}
  H = x J_z - \frac{1 - x}{N} \left(J_+ J_+ + J_- J_-\right),
\end{equation}
where $J_\mu = \sum_{p=1}^N J_{\mu p}$ for $\mu \in \{\pm, z\}$ are the global \SU operators. The parameter $x$ tunes the interaction strength. The system is noninteracting for $x=1$ while correlation strength grows as $x$ is reduced, and becomes extremely correlated at $x=0$. In this paper, we only consider the parameter regime $0\leq x \leq 1$.

The Lipkin model is exactly solvable within the Richardson-Gaudin ansatz.~\cite{ortiz_exactly-solvable_2005, lerma_h_lipkinmeshkovglick_2013}
Exact diagonalization (or full configuration interaction) is also straightforward as the size of the Hilbert space grows linearly with the number of spins. Despite its seemingly simple structure, the Lipkin model exhibits nontrivial physics, particularly near the transition from weakly to strongly correlated regimes; mean-field theory predicts that the parity symmetry, $P = e^{i \pi J_z}$, breaks spontaneously for $x < x_c = (2N - 2)/(3N - 2)$.
For this reason, the Lipkin model serves as an ideal test bed for new computational methods and theories in many-body physics and chemistry (c.f. Refs.~\onlinecite{wahlen-strothman_merging_2017,harsha_difference_2018}).

In an exact theory, spontaneous symmetry breaking occurs in the thermodynamic limit (TDL) but not for finite systems. We can, however, artificially break the symmetry to obtain an energetically lower mean-field solution. This is known as the unrestricted mean-field theory. In our thermal CC implementation, we use an unrestricted mean-field Hamiltonian $H_0$ that allows for ground-state symmetry breaking whenever possible. For Lipkin, this happens for $x \leq x_c$.
Appendix~\ref{app:symmetry-lipkin} contains more details on symmetry breaking and mean-field theory for the Lipkin model.

Finally, it is noteworthy that the Lipkin model can also be considered as system of spinless fermions (see, e.g., Ref.~\onlinecite{wahlen-strothman_merging_2017}). The fermionic Lipkin model has been studied as a benchmark system in thermal cluster cumulant theory, one of the earliest thermal generalizations of CC.~\cite{mandal_finite-temperature_2003} While the spin and fermionic representations come with their own merits and demerits, for the purpose of this article, the \su version merely acts as an exactly solvable benchmark model.

Figure~\ref{fig:lipkin-error} shows the absolute error in the internal energy per site, computed at the level of mean-field and CCSD approximations (Eq.~\ref{lin-resp-exp-val}), as a function of the inverse temperature and for various values of the interaction parameter $x$ in a 32-site Lipkin model. The left (right) panel in the figure presents results for theories constructed from mean-field reference states built with a symmetry-preserving (symmetry-breaking) mean-field Hamiltonian. Thermal CC improves significantly over mean field, particularly for low temperatures. Most of the loss in accuracy occurs at intermediate temperatures and near $x = x_c$, where a single-reference description of the system is inadequate. Both the symmetry-adapted as well as symmetry-broken thermal mean field and CCSD results approach the respective ground-state theories in the limit of zero temperature. The ground-state mean-field (CCSD) energy errors are indicated by the colored ``o'' and ``x'' markers on the $y$-axis in the figure.
The spikes observed in the logarithmic plots for the absolute error in the internal energy occur due to the accidental crossing between the exact and the approximate internal energy curves. This is an effect of the nonvariational nature of the internal energy itself combined with the nonvariational character of the CCSD approximation. Total internal energy results for the 32-site Lipkin model with $x = 0.7$ are presented in Appendix~\ref{app:total-energy-plots}, and shed more light on this issue.
We would also like to mention that for the Lipkin model with $x = 0.5$ (in the right panel), while it may appear otherwise, thermal CC does go to the correct ground-state limit as we evolve the system to a very large value of inverse temperature, $\beta \simeq 1000$.

\subsubsection{Implementation details}
The global \SU symmetry in the Lipkin model allows us to significantly reduce the structure of the cluster operator. We can drop the summation over the lattice indices and use
\begin{align}
  T (\beta) &= t_0 + t_1 S^+ + \tilde{t}_1 \tilde{S}^+ + \alpha_1 Y^{++}
  + m S^+ \tilde{S}^+
  \nonumber
  \\
  & \quad
  + t_2 S^+ S^+ + \tilde{t}_2 \tilde{S}^+ \tilde{S}^+
  + \alpha_2 Y^{++} Y^{++}.
\end{align}
Here, the $su(4)$ operators represent the global operators, i.e., for any generator $X$,
\begin{equation}
  X \equiv \sum_p X_p.
\end{equation}
The operators $Y^{++}$ and $S^{+} \tilde{S}^+$ have overlapping contributions to the thermal state since
\begin{equation}
  S^+ \tilde{S}^+ = \sum_{pq} S^+_p \tilde{S}^+_q,
  \quad
  \textrm{and}
  \quad
  Y^{++} = \sum_p S^+_p \tilde{S}^+_p.
\end{equation}
To avoid making the imaginary-time evolution complicated, we choose either $Y^{++}$ or $S^+ \tilde{S}^+$ (but not both) in our cluster operator. We find that these choices lead to very similar result. Therefore, for all the work presented here, we use $Y^{++}$ but not the $S^+ \tilde{S}^+$ term in the cluster operator.

The number of parameters in the cluster operator is independent of the system size and depends only on the order of approximation. Hence, the computational scaling of thermal CC in Lipkin depends only on the number of grid points over which we perform the integration.
While this implies that we can study large Lipkin models without any added computational cost, we consider a model with merely $32$ spins for the ease of generating exact finite-temperature results.

\subsection{Transverse field Ising model}
The one-dimensional (1D) TFIM is a canonical model to study quantum criticality and phase transitions. The reason behind its popularity is that the model exhibits a quantum phase transition between ordered and disordered phases while also being tractable both analytically and numerically. Consequently, it serves as an ideal model to benchmark new computational theories like our thermal CC. The Hamiltonian is given by
\begin{equation}
  H = - 4 \sum_{i} J^{z}_i J^{z}_{i+1} + 2 g \sum_i J^x_i,
\end{equation}
where $g$ (chosen to be positive) quantifies the strength of the transverse magnetic field. In our work, we only consider 1D chains with periodic boundary condition. In the absence of the transverse field, we have a ferromagnetic Ising model that breaks the $\mathbb{Z}_2$ symmetry in the TDL. On the other hand, for large $g$, the model has a disordered paramagnetic ground state. The one-dimensional chains exhibit a quantum phase transition from a ferromagnetically ordered phase to a disordered phase at $g = 1$. It is well known that the mean-field theory overestimates the magnetic order and predicts a transition at $g = 2$ instead. For our thermal CC theory, once again, we artificially break the $\mathbb{Z}_2$ symmetry to obtain an energetically lower mean-field solution when possible. Appendix~\ref{app:symmetry-tfim} contains further details about the mean-field theory and the choice of $H_0$ for thermal mean field.

\begin{figure}
  \centering
  \includegraphics[width=0.98\linewidth]{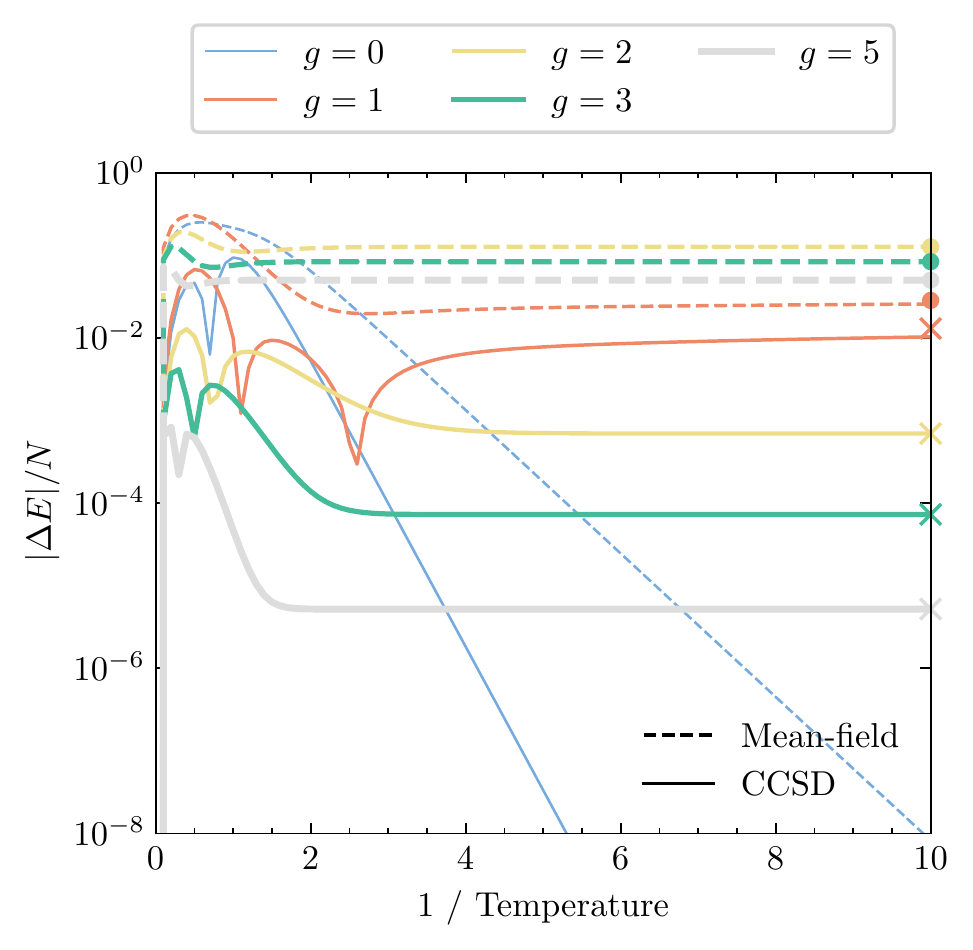}
  \caption{Absolute error in internal energy per site for 10-site transverse field Ising models at various values of the transverse field $g$. The colored ``o'' and ``x'' markers on the $y$-axis indicate the corresponding errors for the ground-state mean-field and CCSD energies, respectively.}
  \label{fig:tfim-error}
\end{figure}

Figure~\ref{fig:tfim-error} plots absolute error in internal energy per site for thermal mean field and CCSD as a function of inverse temperature for a 10-site TFIM at various values of the transverse field. Here, we have used Eq.~\ref{lin-resp-exp-val} to compute the thermal CCSD internal energy. As for Lipkin, thermal CC significantly improves over mean field.
We also observe similar spikes in the log error plots. To corroborate this nonvariational nature of the results, we plot the total mean-field, CCSD and exact internal energies for the 1D TFIM at $g = 1$ in Appendix~\ref{app:total-energy-plots}.
We should note that for $g = 0$ and $g >> 1$, both the mean-field theory and thermal CCSD are exact for the ground state. However, even when ground-state CC or mean-field theory are exact, the finite-temperature theories may not be so. As we can observe from the $g = 0$ results in Fig.~\ref{fig:tfim-error}, both thermal mean field and CC give non-zero errors for the internal energy, which decrease exponentially as we evolve towards zero temperature.

\begin{figure*}
  \centering
  \includegraphics[width=0.85\linewidth]{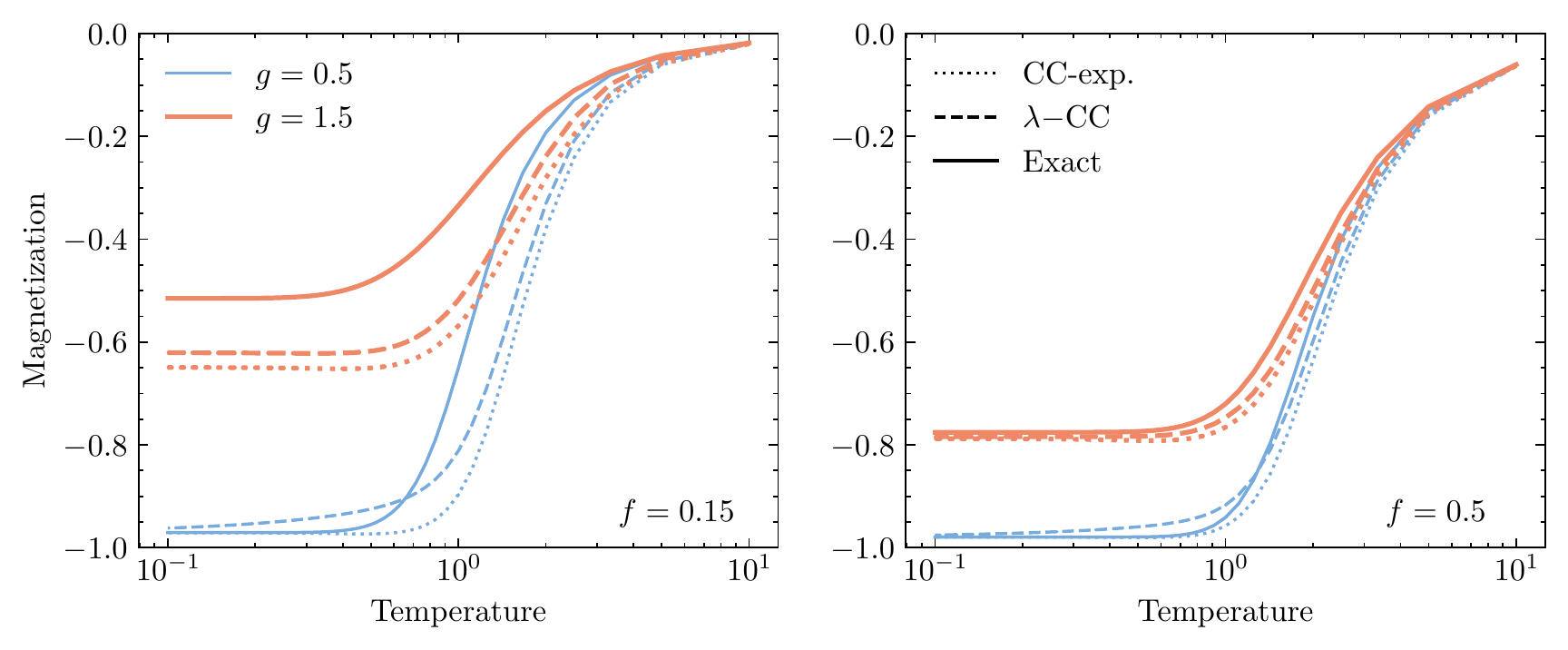}
  \caption{Magnetization curves for 10-site transverse field Ising models for transverse fields $g = 0.5, 1.5$, indicated by red and blue, respectively, and magnetizing fields $f = 0.15$ (left panel) and $f = 0.5$ (right panel). The plots compare magnetization calculated using CC expectation value (Eq.~\ref{lin-resp-exp-val}) and $\lambda$-derivative CC (Eq.~\ref{lambda-exp}) against exact magnetization, which was calculated as the $f$ derivative of the exact free energy. Different colors indicate results for different $g$ values, while the line-styles distinguish between various approximations.}
  \label{fig:tfim-magz}
\end{figure*}

For TFIM, we also compute properties, namely magnetization density and spin-spin correlation functions. We first consider the magnetization density, which we calculate using both the CC expectation value and $\lambda$ derivative of free energy (as already noted in Sec.~\ref{sec:exp-val}). To make a sensible comparison, we first introduce an external field $f$ in the $z$-direction, i.e., we redefine the Hamiltonian as
\begin{equation}
  H = - 4 I \sum_{i} J^{z}_i J^{z}_{i+1} + 2 g \sum_i J^x_i + 2 f \sum_i J^z_i.
  \label{tfim_with_f}
\end{equation}
We have also introduced the Ising coupling constant $I$ for book keeping.
The magnetization density can then be calculated as,
\begin{equation}
  M_z = \lim_{f \rightarrow 0^+} \frac{1}{N} \frac{\partial F}{\partial f}.
  \label{mz-exp-val}
\end{equation}
While this definition holds well in the TDL, there are some caveats when working with finite $N$. Consider the ferromagnetic regime near $g = 0$ and $f > 0$ so that the ground state consists of all spins pointing downwards. Starting from this ground state, we can have two different excitations: single spin flip, for which the excitation energy is $\sim 2I$, and all spin flips i.e., the ferromagnetic state with all spins pointing up. For the latter, the excitation energy is $\sim 2 f N$. For a conventional ferromagnetic phase, single spin flips constitute the low-energy excitations. Therefore, the limit for $f$ in Eq.~\ref{mz-exp-val} should be carefully defined, and we should have $2fN > 2I$, or $f>I/N$. 
Using $I = 1$, magnetization at $g = 0$ should be redefined as
\begin{equation}
  M_z = \lim_{f \rightarrow (1/N)^+} \frac{1}{N} \frac{\partial F}{\partial f}.
\end{equation}
As we go away from $g = 0$, the correct limiting value of the external field for the ferromagnetic phase, $f_c$, will change. To avoid such discrepancies while comparing our benchmark calculations and exact results, we only consider magnetizing fields that are sufficiently large.
Figure~\ref{fig:tfim-magz} shows the temperature dependence of magnetization as we compare the results for CC expectation value (Eq.~\ref{lin-resp-exp-val}) and $\lambda$-derivative approaches (Eq.~\ref{lambda-exp}) against the exact results which were obtained by taking the $\lambda$ derivative of the exact free energy.
In both approximate results, we use a $\lambda$-dependent reference. Since magnetization is a one-body operator, we define the field-dependent thermal reference as
\begin{equation}
  \ket{\Phi (\beta, \lambda)} = e^{-\beta (H_0 + 2 f \sum_i J^z_i) / 2} \ket{\mathbb{I}}.
\end{equation}
While CC expectation values uses a fixed value of the external field (at which the properties are calculated), the $\lambda$-derivative approach considers $f$ dependence of the mean field at every grid point before the derivative is evaluated.
Both the CC expectation value and $\lambda$ derivative perform reasonably well in the weakly correlated regime ($g = 0.5$) for all the values of external field. Near the phase transition region ($g = 1.5$), the overall quality of the CC wave function drops. However, the $\lambda$-derivative results are consistent with the CC expectation values.

\begin{figure*}
  \centering
  \includegraphics[width=0.9\linewidth]{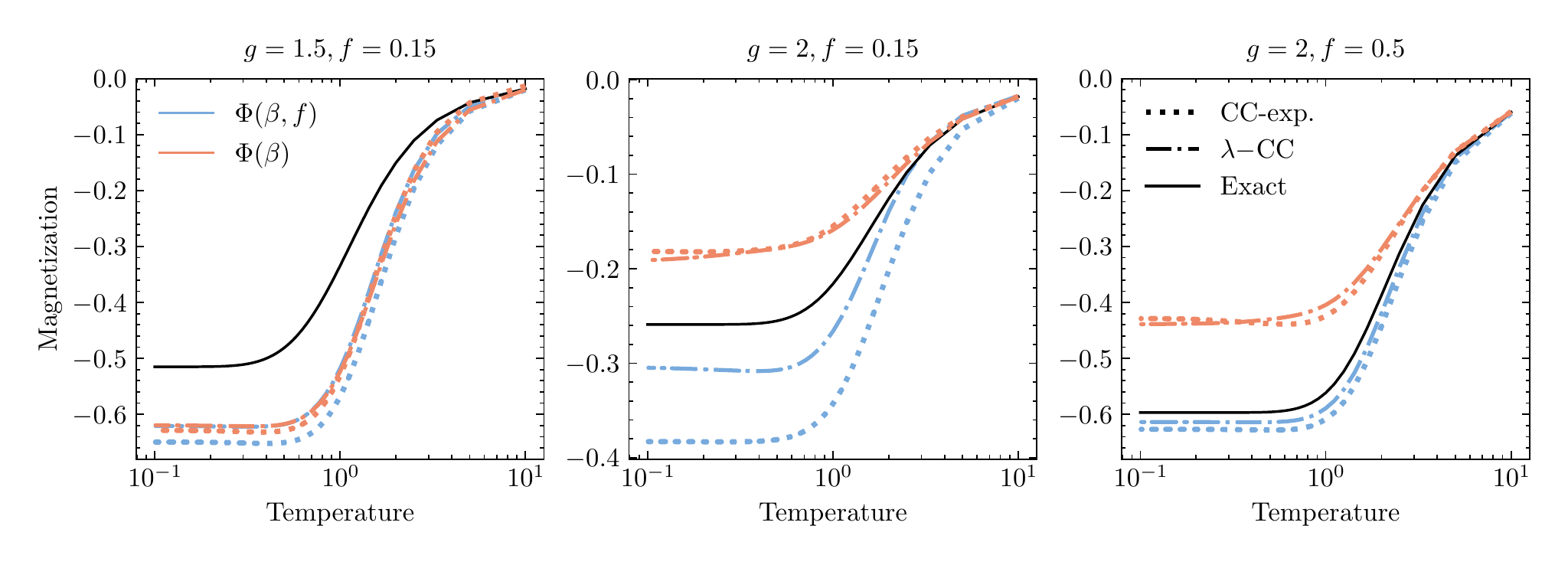}
  \caption{Magnetization densities for 10-site TFIMs with $g = 1.5$, $f=0.15$ in the left, and $g = 2$, $f=0.15$ in the center, and $g = 2$, $f = 0.5$ in the right panels, respectively. We compare CC expecation value and $\lambda$-derivative methods, with and without the external field or $f$ dependence in the mean-field thermal state, against the exact results.}
  \label{fig:compare-lambda-in-mean-field}
\end{figure*}

In order to illustrate the importance of an $f$-dependent mean field, we present magnetization density results for a 10-site TFIM at $(g, f) = (1.5, 0.15)$, $(g, f) = (2, 0.15)$, and $(g, f) = (2, 0.5)$.
Figure~\ref{fig:compare-lambda-in-mean-field} compares CC expectation value and $\lambda$-derivative estimates to the magnetization density, computed with and without an external field or $f$ dependence in the mean-field reference, against exact results. For small external field, as in the case for $(g, f) = (1.5, 0.15)$, expectation values with $f$-dependent and $f$-independent mean-field references are similar, with the latter being marginally better. This is because mean-field theory in 1D TFIM already overestimates the ferromagnetic order. Including the effect of an external field only increases the strength of the magnetization.
On the other hand, as the external field strength is increased, including $f$ dependence in the mean-field theory becomes crucial. This is corroborated by the results for $(g, f) = (2.0, 0.15)$ and $(g, f) = (2.0, 0.5)$.

\begin{figure*}[t]
  \centering
  \includegraphics[width=0.95\linewidth]{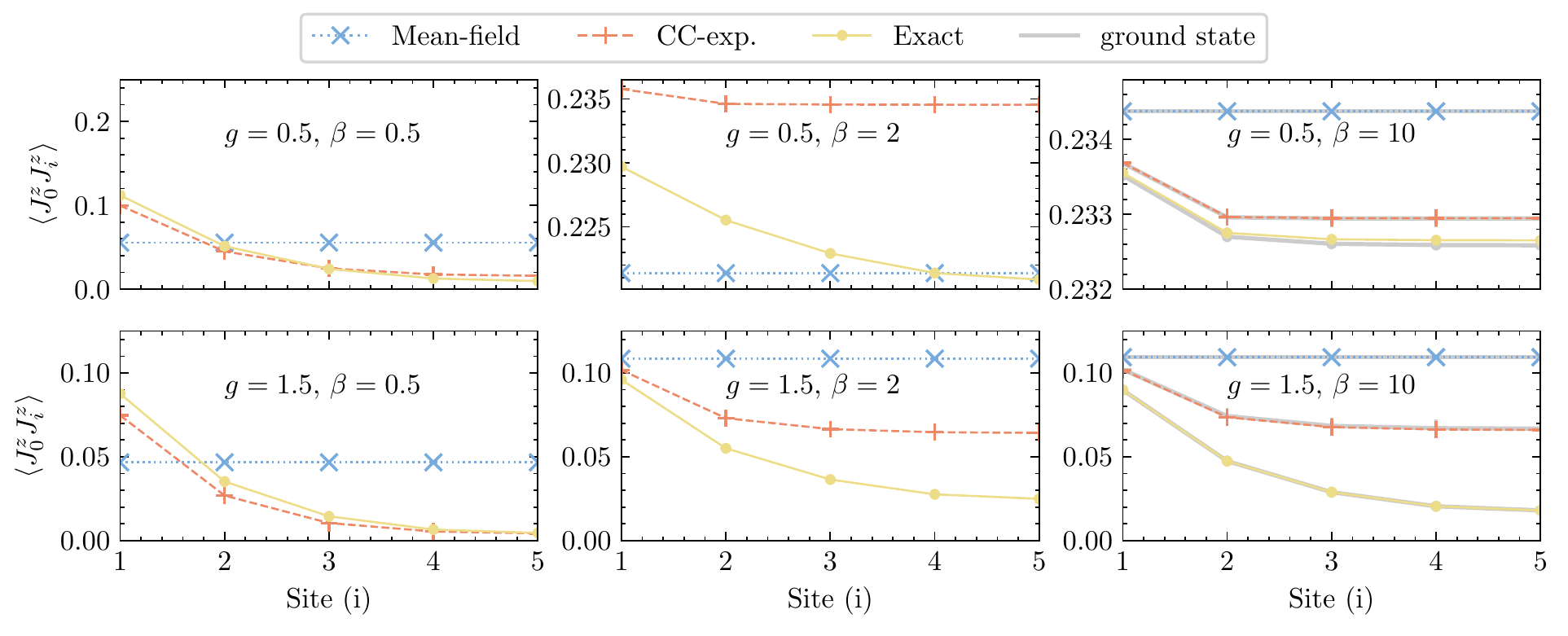}
  \caption{Spin-spin correlation plots for 10-site transverse field Ising models for $g = 0.5$ (top row) and $g = 1.5$ (bottom row) and $f = 0$. We compare the mean-field and CC expectation value (Eq.~\ref{lin-resp-exp-val}) estimates against exact results, which were calculated as ensemble averages, at various values of $\beta$. Results for the corresponding ground-state theories are also plotted in grey in the third column ($\beta = 10$), and demonstrate convergence of thermal theories as $\beta \rightarrow \infty$. For the 10-site model, with periodic boundary conditions, the correlation function $\braket{J^z_0 J^z_i}$ is symmetric about $i = 5$. Therefore we plot data only for $i = 1$ to $5$.}
  \label{fig:tfim-zzcorr}
\end{figure*}

In Fig.~\ref{fig:tfim-zzcorr}, we also compare mean-field and CC expectation value results for the spin-spin correlation function $\braket{J^z_0 J^z_i}$ against exact ensemble averages for this 10-site model. 
Once again, the mean-field approach performs poorly and gives a flat, featureless correlation function. On the other hand, thermal CCSD adds significant corrections, quantitative and qualitative, for both $g = 0.5$ and $g = 1.5$, particularly at low and high temperatures. Near the thermal phase transition region, i.e., at $\beta = 2$, where thermal CCSD is the least accurate, the correlation function, despite exhibiting the right qualitative structure, is not quantitatively accurate (see second column of Fig.~\ref{fig:tfim-zzcorr}).
For large $\beta$, i.e., as we approach zero temperature, the correlation curves converge to the corresponding ground-state properties, as we can deduce from the grey-colored ground-state curves in the third panels of each row in Fig.~\ref{fig:tfim-zzcorr}.
We also observe one of the side effects of the asymmetric expectation values in CC theory. For $g = 0.5$, going from $\beta = 2$ to $\beta = 10$, we find that the strength of thermal CCSD correlation function decreases while the exact correlation increases. While the decrease in correlation is negligible in this case ($\sim 10^{-3}$), the erratic behavior may become severe when CC is not a good approximation to the thermal state.
Higher order approximation to the bra state can be used to address such problems.
Finally, we note that $g = 1.5$ results do not exhibit similar problems.

\subsubsection{Implementation details}
We can exploit the symmetries of the system to simplify the structure of the cluster operator for TFIM, just as we did for the Lipkin model. For the periodic chains under consideration, all sites are equivalent. Therefore, we can express the cluster operator as
\begin{align}
  T(\beta) &= t_0 + s \sum_p S^+_p + \tilde{s} \sum_p \tilde{S}^+_p + \alpha_0 \sum_p Y^{++}_p
  \nonumber
  \\
  & \quad
  + \frac{1}{2} \sum_{pq} \left(
    d_{pq} S^{+}_p S^{+}_q + \tilde{d}_{pq} \tilde{S}^+_p \tilde{S}^+_q + m_{pq} Y^{++}_p Y^{++}_q
  \right)
  \nonumber
  \\
  & \quad
  + \sum_{pq} x_{pq} S^{+}_p \tilde{S}^{+}_q,
\end{align}
where we enforce that the tensors $d$, $\tilde{d}$, $m$ and $x$ are symmetric and have zero diagonals. With these simplifications, we can perform the imaginary-time evolution with a computational scaling of $\mathcal{O}(N^4 N_{grid})$. This scaling can be further brought down by realizing that the tensors $d_{pq}$, $\tilde{d}_{pq}$, $m_{pq}$ and $x_{pq}$ depend only on $|p - q|$.

\section{Conclusion}
We have demonstrated a generalization of coupled cluster theory to finite-temperature \SU systems with the help of thermofield dynamics. The CCSD framework, which has been applied successfully to study ground-state properties of spin systems, performs equally well at finite temperatures. At the same time, thermal CC experiences the same problems as its ground-state counterpart: it is insufficient for strongly correlated systems, and the expectation values are asymmetric. The single-reference nature of the CC ansatz fails to capture the underlying multi-configurational physics in strongly correlated systems. This, in conjunction with the fact that CC is a nonvariational theory, leads to poor performance, both for the internal energy and properties.
However, a key advantage of coupled cluster is that it is systematically improvable. Including higher order excitations (triples, quadruples, etc.) in both the cluster operator and the configuration interaction approximation to the bra state can help alleviate some of the associated issues. In fact, for spin lattices, where CCSD scales as $\mathcal{O}(N^3)$ or $\mathcal{O}(N^4)$, higher order excitations can be added without making the theory computationally intractable.
Another alternative is to use a multi-configurational wave functions (such as nonorthogonal CI) or unconventional mean-field states (such as spin antisymmetrized geminal power state, dimer mean field, resonating valence bond, etc.) as reference states to build the thermal CC ansatz. Such strategies have been explored in the context of ground-state CC theory and have shown promise. Their thermal generalizations provide pathways to expand upon our current work.

\begin{acknowledgments}
This work was supported by the U.S. Department of Energy, Office of Basic Energy Sciences, Computational and Theoretical Chemistry Program under Award No. DE-FG02-09ER16053. G.E.S. acknowledges support as a Welch Foundation Chair (Grant No. C-0036).
Y. X. acknowledges Zhejiang University and Rice University for sponsoring his summer internship at Rice, during which he made contributions to this work.
\end{acknowledgments}

\appendix

\section{\label{app:symmetry-lipkin}Symmetry breaking in Lipkin model}
The Lipkin-Meshkov-Glick Hamiltonian, given by
\begin{equation}
  H = x J_z - \frac{1 - x}{n} \big(J_+J_+ + J_-J_-\big),
\end{equation}
commutes with the Casimir operator,
\begin{equation}
  J^2 = \frac{J_+ J_- + J_- J_+}{2} + J_z J_z.
\end{equation}
The $J^2$ eigenvalue is determined by the total number of spins; for $n$ spins, the eigenstates of $H$ reside in the $j=n/2$ sector.
The Hamiltonian also commutes with the parity operator, $\hat{P} = e^{i \pi J_z}$, the symmetry of interest. For a given spin configuration, the parity eigenvalue is $1$ (or $-1$) if the difference in the number of up- and down-spins is even (or odd). Therefore, simultaneous eigenstates of $H$ and $\hat{P}$ are made out of spin configurations, all of which have only even or only odd parity. As a result, the symmetry-preserving mean field [or the restricted Hartree-Fock (RHF)] state that optimizes the energy is simply the configuration with all down-spins, i.e.,
\begin{equation}
  \ket{\phi_{RHF}} = \bigotimes_i \ket{\dn}_i.
\end{equation}
The corresponding mean-field Hamiltonian and the mean-field ground-state energy are
\begin{subequations}
  \begin{align}
    H_0^{RHF} &= x J_z,
    \\
    E_{RHF} &= -\frac{n x}{2}.
  \end{align}
\end{subequations}
While working in the symmetry adapted basis, we use $H_0^{RHF}$ to define the mean-field thermal state.

We can also consider a rotated product state, that is not an eigenvector of the parity operator, as the ground-state mean-field reference. This unrestricted Hartree-Fock (UHF) state can be expressed as
\begin{equation}
  \ket{\phi_{UHF}} = \frac{1}{(1 + \kappa^2)^{n/2}} e^{\kappa J_+} \ket{\phi_{RHF}},
\end{equation}
where $\kappa$ parametrizes the UHF state. Using a UHF state as an approximation to the ground state is equivalent to a rotation of the underlying \su algebra such that the UHF wave function has all down-spins in the new basis. The \su operators in the new basis are related to the original operators via the following transformation,
\begin{subequations}
  \begin{align}
    J_+ &= \frac{K_+ - \kappa^2 K_- - 2\kappa K_z}{1 + \kappa^2},
    \\
    J_- &= \frac{K_- - \kappa^2 K_+ - 2\kappa K_z}{1 + \kappa^2},
    \\
    J_z &= \frac{\kappa (K_+ + K_-) + (1 - \kappa^2) K_z}{1 + \kappa^2}.
  \end{align}
\end{subequations}
The Hamiltonian, expressed in terms of the $K$ operators, becomes
\begin{align}
  H &= h_z K_z + h_{\pm} (K_+ + K_-) + v_{\pm} (K_+K_+ + K_-K_-)
  \nonumber
  \\
  & \quad + v_z K_z^2 + v_{\times} K_+ K_- + v_{\pm z} (K_+ K_z + K_z K_-).
\end{align}
Exact expressions for the Hamiltonian parameters in the above expression can be found in the appendix of Ref.~\onlinecite{wahlen-strothman_merging_2017}. The UHF mean-field Hamiltonian and the corresponding mean-field energy are
\begin{subequations}
  \begin{align}
    H_0^{UHF} &= h_z K_z - \frac{v_z n}{2} K_z,
    \\
    E_{UHF} &= -\frac{h_z n}{2} + \frac{v_z n^2}{4}.
  \end{align}
\end{subequations}
The UHF energy is minimized to find the optimal value of the rotation parameter $\kappa$. As we have mentioned in the main text, $\kappa \neq 0$ only for $x \leq x_c$. While working in the broken-symmetry regime (i.e., $x \leq x_c$), we use $H_0^{UHF}$ to define the mean-field thermal state.

\section{\label{app:symmetry-tfim}Symmetry breaking in TFIM}
The TFIM has a $\mathbb{Z}_2$ symmetry, i.e., if all the spins (in $z$-basis) are flipped, then the energy remains unchanged. In the Ising limit, i.e., for $g = 0$, the model has a doubly-degenerate ferromagnetic ground-state. In the thermodynamic limit, the $\mathbb{Z}_2$ symmetry breaks spontaneously. For finite systems, we can break this symmetry artificially by introducing the following symmetry broken mean-field state to approximate the ground state,
\begin{equation}
  \ket{\phi_{UHF}} = \bigotimes_p \left(-\sin \frac{\theta}{2} \ket{\up}_p + \cos \frac{\theta}{2} \ket{\dn}_p \right).
\end{equation}
The mean-field Hamiltonian and its corresponding energy depend on the rotation parameter $\theta$, and for the one-dimensional periodic case with $N$ spins, in the presence of an external magnetizing $z$-field $f$ (see Eq.~\ref{tfim_with_f} for the Hamiltonian), they are given by
\begin{subequations}
  \begin{align}
    H_0 &= 2 \sum_i \left((\cos \theta + f) J^z_i + g J^x_i\right),
    \\
    E_{UHF} &= -N \big(
      \cos^2 \theta + g \sin \theta + f \cos \theta
    \big).
  \end{align}
\end{subequations}
The UHF energy is minimized with respect to the rotation parameter $\theta$, the optimal values of which (at $f = 0$) are
\begin{equation}
  \theta = \begin{cases}
    \arcsin g / 2, & \quad g \leq 2
    \\
    \pi / 2, & \quad g > 2
  \end{cases}.
\end{equation}
We use this $H_0$ (with optimized $\theta$) that breaks the $\mathbb{Z}_2$ symmetry, except when $\theta = \pi / 2$, to construct the mean-field thermal state for the TFIM.

In the same way as for the Lipkin model, we can introduce a rotation in the \su algebra so that in the new basis, the UHF state corresponds to all down-spins. The full Hamiltonian in the rotated basis then becomes
\begin{align}
  H &= -4 \sum_{i} \Big[
    \cos^2 \theta K^z_i K^z_{i+1} + \sin^2 \theta K^x_i K^x_{i+1} 
    \nonumber
    \\
    & 
    \quad - \frac{1}{2}\sin 2\theta \Big(K^x_i K^z_{i+1} + K^z_i K^x_{i+1}\Big)
    \nonumber
    \\
    &
    \quad + \frac{g}{2} \Big(\cos \theta K^x_i + \sin \theta K^z_i\Big)
  \Big].
\end{align}
Similarly, the mean-field Hamiltonian, written in the new basis, becomes simply
\begin{equation}
  H_0 = -2 \sum_i \Big(
    -\cos^2 \theta + g \sin \theta
  \Big) K^z_i.
\end{equation}
While the results do not depend on the choice of the basis, it is generally convenient to construct CC-like wave functions starting with a mean-field state that corresponds to all down-spins.
Our description of the mean-field theory for the TFIM closely follows Ref.~\onlinecite{rosenfeld_phase_2000} and we recommend the reader to refer to this article for further details.

\section{\label{app:total-energy-plots}Total internal energy for 1D TFIM}
\begin{figure*}[t]
  \centering
  \includegraphics[width=0.4\linewidth]{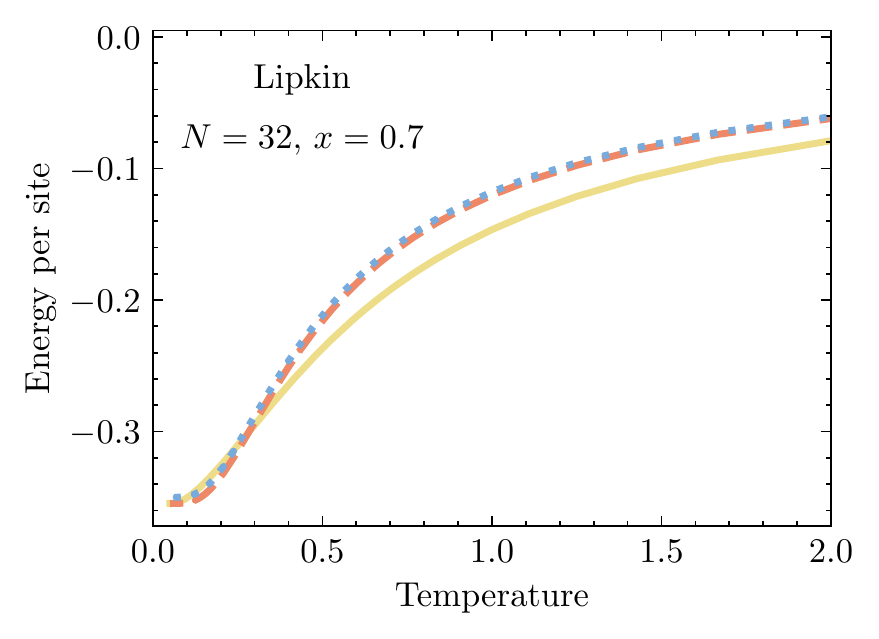}
  \includegraphics[width=0.4\linewidth]{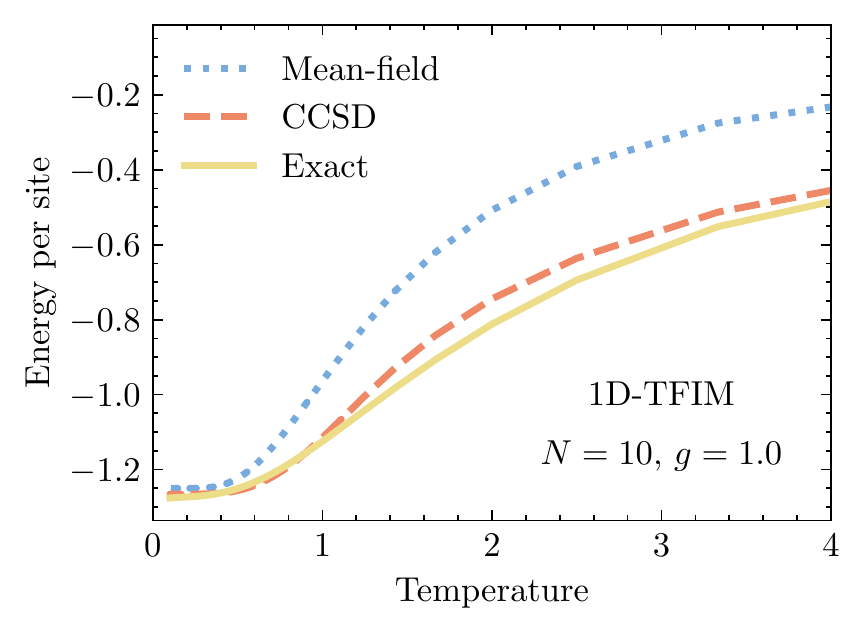}
  \caption{Total internal energy per site for (left) 32-site Lipkin model with $x = 0.7$, and (right) 10-site transverse field Ising model at $g = 1.0$. The figure compares mean-field and CCSD approximations against the exact results.}
  \label{fig:total-energy-plots}
\end{figure*}

The logarithmic plots for absolute internal energy errors (Figs.~\ref{fig:lipkin-error} and \ref{fig:tfim-error}) show several spikes that correspond to zero error. These occur when the approximate and exact internal energy curves intersect. These are a result of the nonvariational nature of the internal energy and may be worsened by the nonvariational character of the CC theory.
In Fig.~\ref{fig:total-energy-plots}, we plot the total internal energy per site for the 32-site Lipkin model with $x = 0.7$ in the left panel, and the 10-site 1D TFIM at $g = 1$ in the right panel.
It is clear that neither the mean-field nor the CCSD internal energies provide a variational upper bound to the exact internal energy.
For example, in the 1D TFIM plot, the CCSD internal energy crosses the exact curve at two different temperatures. These correspond to the two spikes that are observed in Fig.~\ref{fig:tfim-error}.

\section{\label{app:thermal-properties}Thermal Properties}
In Sec.~\ref{sec:exp-val}, we described two different ways to compute thermal properties: CC expectation values and $\lambda$ derivatives.
Here, we will discuss a few details about the equivalence, or lack thereof, between the two approaches.

\subsection{Exact theory}
For a first-derivative property (e.g. the dipole moment or magnetization density), the property can be computed as the derivative of the free energy. We thus define a $\lambda$-dependent thermal state,
\begin{equation}
  \ket{\Psi (\beta, \lambda)} =
  e^{-\beta H(\lambda) / 2} \ket{\mathbb{I}},
\end{equation}
where $H(\lambda) = H + \lambda \mathcal{O}$. The norm of $\ket{\Psi (\beta, \lambda)}$ gives us the partition function, $\mathcal{Z}(\beta, \lambda) = \braket{\Psi (\beta, \lambda) | \Psi (\beta, \lambda)}$, from which we can extract the free energy and write the thermal average of $\mathcal{O}$ as
\begin{equation}
  \braket{\mathcal{O}} = -\frac{1}{\beta} \frac{\partial}{\partial \lambda} \ln \mathcal{Z} (\beta, \lambda).
  \label{eq:app:thermal-properties-lambda-exp}
\end{equation}
The $\lambda$ derivative of the partition function can be simplified using the Wilcox identity,~\cite{wilcox_exponential_1967}
\begin{subequations}
  \begin{align}
    \frac{
      \partial \mathcal{Z}(\beta, \lambda)
    }{\partial \lambda}
    &= \frac{\partial}{\partial \lambda}
    \braket{
      \mathbb{I} |
      e^{-\beta H(\lambda)} |
      \mathbb{I}
    },
    \\
    &= -\beta \int_0^1 d\alpha \braket{
      \mathbb{I} |
      e^{-(1 - \alpha) \beta H(\lambda)} 
      \mathcal{O} e^{-\alpha \beta H(\lambda)} |
      \mathbb{I}
    }.
  \end{align}
\end{subequations}
Using the fact that for a physical operator $X$, $\braket{\mathbb{I} | X | \mathbb{I}} = \mathrm{Tr} X$, combined with the cyclic property of trace, we get
\begin{subequations}
  \begin{align}
    \frac{
      \partial \mathcal{Z}(\beta, \lambda)
    }{\partial \lambda}
    &=
    -\beta \int_0^1 d\alpha
    \mathrm{Tr} \left(
      e^{-(1-\alpha) \beta H(\lambda)} \mathcal{O} e^{-\alpha \beta H(\lambda)}
    \right),
    \\
    &=
    -\beta \int_0^1 d\alpha
    \mathrm{Tr} \left(
      e^{-\beta H(\lambda) / 2} \mathcal{O} e^{-\beta H(\lambda) / 2}
    \right),
    \\
    &= -\beta \braket{
      \Psi (\beta, \lambda) | \mathcal{O} | \Psi (\beta, \lambda)
    }.
  \end{align}
\end{subequations}
Therefore, in the exact theory, the $\lambda$ derivative of the free energy is equivalent to the expectation value of $\mathcal{O}$ over the thermal state, i.e.,
\begin{equation}
  \braket{\mathcal{O}} = -\frac{1}{\beta} \frac{\partial}{\partial \lambda} \ln \mathcal{Z} (\beta, \lambda)
  = \frac{
    \braket{\Psi (\beta, \lambda) | \mathcal{O} | \Psi (\beta, \lambda)}
  }{
    \braket{\Psi (\beta, \lambda) | \Psi (\beta, \lambda)}
  }.
\end{equation}

\subsection{Mean-field theory}
Unlike the exact theory, the expectation value and the free-energy derivative do not necessarily lead to the same results in the mean-field theory.
The expectation value of a physical operator $\mathcal{O}$ over the mean-field thermal state is defined as
\begin{equation}
  \braket{\mathcal{O}}_{0, \textrm{exp.}}
  = \braket{0(\beta) | \mathcal{O} | 0(\beta)}.
  \label{eq:app:thermal-properties-mean-field-expectation}
\end{equation}
On the other hand, we can also define $\braket{\mathcal{O}}_0$ as the $\lambda$ derivative of the mean-field free energy, i.e.,
\begin{equation}
  \braket{\mathcal{O}}_{0, \lambda} = -\frac{1}{\beta} \lim_{\lambda \rightarrow 0}
  \frac{\partial}{\partial \lambda} \ln \braket{\Phi (\beta, \lambda) | \Phi (\beta, \lambda)},
  \label{eq:app:thermal-properties-mean-field-lambda}
\end{equation}
where 
\begin{equation}
  \ket{\Phi (\beta, \lambda)} = e^{-\beta (H_0 + \lambda \mathcal{O}_0) / 2} \ket{\mathbb{I}},
\end{equation}
with $\mathcal{O}_0$ being the mean-field (or one-body) contribution to $\mathcal{O}$.
For a general two-body $\mathcal{O}$, Eqs.~\ref{eq:app:thermal-properties-mean-field-expectation} and \ref{eq:app:thermal-properties-mean-field-lambda} are not equivalent.
The thermal expectation value of only the mean-field component of $\mathcal{O}$ is identical to the $\lambda$ derivative, i.e.,
\begin{equation}
  \braket{\mathcal{O}}_{0, \lambda} = \braket{\mathcal{O}_0}_{0, \textrm{exp.}},
  \label{eq:app:thermal-properties-equiv-mf-linres-lambda}
\end{equation}
which implies that only when the observable $\mathcal{O}$ is a one-body operator, the mean-field expectation value and $\lambda$ derivative of the mean-field free energy are identical.

Note that in the foregoing equations, we have ignored the orbital optimization of the mean-field free energy in the presence of $\lambda \mathcal{O}_0$, which, if included, will result into a mean-field Hamiltonian $H_0(\beta, \lambda)$ that is nonlinear in $\lambda$.

\subsection{Coupled cluster theory}
It follows directly from mean-field theory that for approximate wave function theories, expectation values over thermal state and free-energy derivatives do not necessarily produce identical results. Let us see an explicit proof for the coupled cluster theory.
The CC expectation value and $\lambda$-derivative expressions are given in Eqs.~\ref{lin-resp-exp-val} and \ref{lambda-exp},
\begin{subequations}
  \begin{align}
    \braket{\mathcal{O}}_{CC,\, \textrm{exp.}}
    &= \frac{
      \braket{\Psi_L | \mathcal{O} | \Psi_R}
    }{
      \braket{\Psi_L | \Psi_R}
    },
    \\
    \braket{\mathcal{O}}_{CC,\, \lambda}
    &= \lim_{\lambda \rightarrow 0} \frac{\partial F}{\partial \lambda},
  \end{align}
\end{subequations}
where $F$ is the free energy of the system, and is defined as
\begin{subequations}
  \begin{align}
    F (\beta, \lambda) &= -\frac{1}{\beta} \ln \mathcal{Z}_{CC} (\beta, \lambda),
    \\
    &= \frac{1}{\beta} \int_0^\beta d \tau E_{CC} (\tau, \lambda),
  \end{align}
\end{subequations}
with $E_{CC} (\tau, \lambda) = \braket{\Psi_L | H(\lambda) | \Psi_R} / \braket{\Psi_L | \Psi_R}$.
Recall that $\bra{\Psi_L}$, $\ket{\Psi_R}$, and the CC partition function $\mathcal{Z}_{CC}$ are given by
\begin{subequations}
  \begin{align}
    \bra{\Psi_L} &=
    \bra{\Phi (\beta, \lambda)} (1 + Z) e^{-T} e^{z_0},
    \\
    \ket{\Psi_R} &=
    e^{t_0 + T} \ket{\Phi (\beta, \lambda)},
    \\
    \mathcal{Z}_{CC} (\beta, \lambda)
    &= \braket{\Phi (\beta, \lambda) | \Phi (\beta, \lambda)} e^{t_0 + z_0}.
  \end{align}
\end{subequations}
To show that CC expectation value and $\lambda$-derivative approaches are different, we will show that
\begin{equation}
  \frac{\partial}{\partial \beta} \left(
    \beta \braket{\mathcal{O}}_{CC,\, \textrm{exp.}}
  \right)
  \neq
  \frac{\partial}{\partial \beta} \left(
    \beta \braket{\mathcal{O}}_{CC,\, \lambda}
  \right).
\end{equation}
The left-hand side (LHS) can be simplified as:
\begin{widetext}
  \begin{subequations}
    \begin{align}
      \textrm{LHS}
      &= \left(
        1 + \beta \frac{\partial}{\partial \beta}
      \right) \frac{
        \braket{\Psi_L | \mathcal{O} | \Psi_R}
      }{
        \braket{\Psi_L | \Psi_R}
      },
      \\
      &= \braket{\mathcal{O}}_{CC,\, \textrm{exp.}}
      + \frac{\beta}{2} \left(
        2 E(\beta) \braket{\mathcal{O}}_{CC,\, \textrm{exp.}}
        - \frac{
          \braket{\Psi_L | (H \mathcal{O} + \mathcal{O} H) | \Psi_R}
        }{
          \braket{\Psi_L | \Psi_R}
        }
      \right)
    \end{align}
  \end{subequations}
  On the other hand, the right-hand side (RHS), before taking the limit of $\lambda \rightarrow 0$, simplifies into
  \begin{subequations}
    \begin{align}
      \textrm{RHS}
      &= \frac{\partial}{\partial \beta}
      \int_0^\beta d\tau \frac{
        \partial E_{CC} (\tau, \lambda)
      }{
        \partial \lambda
      },
      \\
      &= \frac{
        \partial E (\beta, \lambda)
      }{
        \partial \lambda
      }
      = \frac{\partial}{\partial \lambda}
      \braket{
        0 (\beta, \lambda) |
        (1 + Z) e^{-T} H (\lambda) e^T |
        0 (\beta, \lambda)
      }
      \\
      &= \braket{\mathcal{O}}_{CC,\, \textrm{exp.}}
      + \braket{
        0 (\beta, \lambda) |
        \frac{\partial Z}{\partial \lambda} e^{-T} H (\lambda) e^T |
        0 (\beta, \lambda)
      }
      \nonumber
      \\
      &
      \quad + \braket{
        0 (\beta, \lambda) |
        (1 + Z) \left(
          \frac{
            \partial e^{-T}
          }{
            \partial \lambda
          } H e^T
          + e^{-T} H \frac{
            \partial e^T
          }{
            \partial \lambda
          }
        \right) |
        0 (\beta, \lambda)
      }
      \nonumber
      \\
      & \quad + \frac{
        \partial \bra{0 (\beta, \lambda)}
      }{
        \partial \lambda
      }
      (1 + Z) e^{-T} H e^T \ket{0 (\beta, \lambda)}
      + \bra{0 (\beta, \lambda)} (1 + Z) e^{-T} H e^T \frac{
        \partial \ket{0 (\beta, \lambda)}
      }{\partial \lambda}.
    \end{align}
  \end{subequations}
  At a first glance, the left- and right-hand side expressions do not show any resemblance beyond the $\braket{\mathcal{O}}_{CC,\, \textrm{exp.}}$ term. Even if we ignore the $\lambda$ dependence in the mean-field reference, and therefore the quasiparticle excitation and deexcitation operators, we get
  \begin{align}
    \textrm{RHS}
    &= \braket{\mathcal{O}}_{CC,\, \textrm{exp.}}
    + \sum_\mu \frac{\partial z_\mu}{\partial \lambda} \braket{
      0 (\beta) |
      \tau_\mu^\dagger e^{-T} H (\lambda) e^T |
      0 (\beta)
    }
    \nonumber
    \\
    &
    \quad + \sum_\mu \frac{\partial t_\mu}{\partial \lambda} \braket{
      0 (\beta) |
      (1 + Z) \left[e^{-T} H(\lambda) e^T, \tau_\mu \right] |
      0 (\beta)
    },
  \end{align}
\end{widetext}
which is clearly different from LHS.
It is well established in ground-state CC theory that, in the absence of orbital relaxation effects, i.e., using a $\lambda$-independent mean-field reference, CC expectation values and $\lambda$-derivative approach results in identical results.
However, the same is not obviously true at nonzero temperatures.

\bibliography{spin-thermo}

\end{document}